\documentclass[longbibliography,preprint,onecolumn]{revtex4-1}

\usepackage{amssymb,amsfonts,amsmath}
\usepackage{graphicx}
\usepackage{color}

\begin{document}


\title{Thermodynamics of firms' growth}

\author{Eduardo Zambrano}
\email{zambrano@pks.mpg.de}
\affiliation{Max-Planck-Institut f\"ur Physik komplexer Systeme, N\"othnitzer Str. 38, 01187 Dresden, Germany
}
\author{Alberto Hernando}
\email{alberto.hernandodecastro@epfl.ch}
\affiliation{SThAR, EPFL Innovation Park, B\^atiment C, 1015 Lausanne, Switzerland}
\affiliation{Laboratory of Theoretical Physical Chemistry, Institut des Sciences et
Ing\'{e}nierie Chimiques, \'{E}cole Polytechnique F\'{e}d\'{e}rale de Lausanne
(EPFL), CH-1015 Lausanne, Switzerland}
\author{Aurelio Fern\'andez-Bariviera}
\affiliation{Department of Business, Universitat Rovira i
Virgili, Av. Universitat 1, 43204 Reus, Spain}
\author{Ricardo Hernando}
\affiliation{SThAR, EPFL Innovation Park, B\^atiment C, 1015 Lausanne, Switzerland}
\author{Angelo Plastino}
\affiliation{National University of La Plata, Physics
Institute (IFLP-CCT-CONICET) C.C.737, 1900 La Plata, Argentina}

\begin{abstract}
The distribution of firms' growth and firms' sizes is  a topic under intense scrutiny.
In this paper we show that a thermodynamic model based on the
Maximum Entropy Principle, with dynamical prior information, can be
constructed that adequately describes the
dynamics and distribution of firms' growth. Our theoretical framework is tested
against a comprehensive data-base of Spanish firms, which covers
to a very large extent Spain's economic activity with a total of 
$1\,155\,142$ firms evolving along a full decade. We show that
the empirical exponent of Pareto's law, a rule often observed in the rank distribution
of large-size firms, is explained by the capacity of the economic system
for creating/destroying firms, and can be used to measure
the health of a capitalist-based economy. Indeed, our model predicts that
when the exponent is larger that 1, creation of firms is favored;
when it is smaller that 1, destruction of firms is favored instead; and
when it equals  1 (matching Zipf's law), the system is in a full macroeconomic
equilibrium, entailing ``free''  creation and/or destruction of firms.
For medium and smaller firm-sizes, the dynamical regime changes;
the whole distribution can no longer  be fitted to a single
simple analytic form and numerical prediction
is required. Our model constitutes the basis of a full predictive framework for 
the economic evolution of an ensemble of firms that can be potentially used
to develop simulations and test hypothetical scenarios, as economic
crisis or the response to specific policy measures.
\end{abstract}

\maketitle


\section{Introduction\label{sec:intro}} 

Many natural, social and economic phenomena follow power
laws. It has been previously ascertained  that the distribution of
incomes \cite{Pareto}, size of cities
\cite{Krugman,Gabaix1999,Bettencourt,Hernando2012,Hernando2013c,Hernando2013,Hernando2013a,Hernando2015},
evolution of human language \cite{Cancho}, internet and genetic
networks \cite{Barabasi}, and scientific publications and
citations \cite{Katz,vanRaan}, all follow power laws. Finding a
complete theory for describing these kind of systems seem an
impractical task, given the huge amount of degrees of freedom
involved these social systems. Notwithstanding, remarkable
regularities were reported and studied, such as Zipf's law
\cite{Zipf,Newman,Hernando2009,Batty}, or the celebrated Gibrat's
Law of proportional growth \cite{Gibrat1931}, which constitutes important milestones
on the quest for a unified framework that mathematically describe
predictable tendencies.

Firm size distributions (FSD) are the outcome of the complex
interaction among several economic forces. Entry of new firms,
growth rates, business environment, government regulations, etc.,
may shape different FSD. The underlying dynamics that drives the
distribution of firms' sizes is still an issue under intense
scrutiny. According to Gaffeo et al. \cite{Gaffeo2012}, there is an
active debate going on among industrial organization's scholars, in which
log-normal, Pareto, Weibull, or a mixture of them, compete for the
best-fitting distributions of FSD. One of the controversial issues
is the very definition of ``size'', which can be measured by
different proxies such as annual sales, number of employees, total
assets, etc.

The seminal contribution by Gibrat \cite{Gibrat1931} initiated a
research line concerning  the formal model that governs firms' sizes
and industry structure. The introduction of a theoretical model
that would underlie the industrial demography could
 be of great help for authorities interested in maintaining fair  competence and/or antitrust policies.

Hart and Prais \cite{HartPrais56} find, using a database of large
firms, that average growth rates and sizes are independent
variables. Quandt \cite{Quandt1966} states that Pareto's
distribution is often rejected when analyzing industries
sub-sectors.
 Other independent empirical studies, carried out by Simon and
Bonnini \cite{SimonBonnini}, Mansfield \cite{Mansfield},
and Bottazzi and Secchi \cite{Bottazzi},
among others, confirm that firms' growth rates are not related to
firm size and that FSD follow a log-normal distribution. Jacquemin
and Cardon de Lichtbuer \cite{Jacquemin} study the degree of firms
and industry concentration in British firms using Fortune's 200
largest industrial companies outside the United States, ranked
according to sales. This study detects an increasing degree of
concentration.

Kwasnicki \cite{Kwasnicki} affirms that skewed size distributions
could be found even in the absence of economies of scale, and that
the shape of the distribution is the outcome of innovation in
firms. In particular, according to his simulations, cost improving
innovations generate Pareto-like skewed distributions. This work
also reconciles the finding by Ijiri and Simon \cite{IjiriSimon}
about the concavity toward the origin of the log-log rank size
plot. Such concavity could be produced by the evolutionary forces
and innovation in the market. Jovanic \cite{Jovanic} finds that
rates of growth for smaller firms are larger and more variable
than those of bigger firms. Similar results are found empirically
for Dutch companies by Marsili \cite{Marsili2005}. On contrary,
Vining \cite{Vining} had argued that the origin of the concavity
is the existence of decreasing returns to scale.

Segal and Spivak \cite{Segal} develop a theoretical model in
which, under the presence of bankruptcy costs, the rate of growth
of small firms is prone to be higher and more variable than that
of larger firms. The same model also predicts that, for the largest
firms, the sequence of growth rates is convergent satisfying 
Gibrat's law, namely
\begin{equation}
\label{eq:eq1}
\dot{x}_i(t) = v_{i}(t) x_i(t),
\end{equation}
where $x_i(t)$ is the size of the $i$th firm at time $t$, $\dot{x}_i(t)$
its change in time, and $v_{i}(t)$ a size-independent growth rate.
This model is consistent with some previous
empirical evidence, as that of Mansfield \cite{Mansfield}.

Sutton \cite{Sutton1997} has published a review of the
literature on markets' structure, highlighting the current
challenges concerning FSD modelling.

During the 1990s, the interest in FSD experienced a revamp, with the
availability of new data-bases. A drawback of early studies was
a biased selection of firms. Typically, data comprise only
publicly traded firms, i.e., the largest ones. In recent years,
new, more comprehensive data sources became available.

Stanley et al. \cite{Stanley1995}, use the Zipf-plot technique in
order to verify fittings of selected data for US manufacturing
firms and find a non-lognormal right tail. Shortly afterwards,
Stanley et al. \cite{Stanley1996} encounter that the distribution
of growth rates has an exponential form. Kattuman \cite{Kattuman}
studies intra-enterprise business size distributions, finding also
a skewed distribution. Axtell \cite{Axtell2001}, using Census data
for all US firms, encounters that the FSD is right skewed, giving
support for the workings of Pareto's law. A similar finding is
due to Cabral and Mata \cite{Cabral2003} for Portuguese
manufacturing firms, although a log-normal distribution
underestimates the skewness of the distribution and is not
suitable for its lower tail. In this line, Fu et al. \cite{Fu2005}
find that, for pharmaceutical firms in 21 countries, and for US
publicly traded firms, growth rates exhibit a central-portion
distributed according to  a Laplace distribution, with power law
tails. Palestrini \cite{Palestrini2007} agrees with a power law
distribution for firm sizes, although he models firm growth as a
Laplace distribution, that could change over business cycles.

According to Riccaboni et al. \cite{Riccaboni2008}, the
simultaneous study of firm sizes and growth presents an intrinsic
difficulty, arising from two  facts: (i) the size distribution follows
a Pareto Law and (ii) firms' growth rate is independent of the
firm's size. This latter property is known as the ``Law of
Proportionate Effect''. Growiec et al. \cite{Growiec2008} study
firms' growth and size distributions using firms' business units
as units of measurements. This study reveals that the size of
products follow a log-normal distribution, whereas firm-sizes
decay as a power law.

Gaffeo et al. \cite{Gaffeo2012}, using data from 38 European
countries, find that log mean and log variance size are linearly
related at sectoral levels, and that the strength of this
relationship varies among countries. Di Giovanni et al.
\cite{diGiovanni} find that the exponent of the power law for
French exporting firms is lower than for non-exporting firms,
raising the argument of the influence of firm heterogeneity in the
industrial demography. Additionally, Gallegati and Palestrini
\cite{Gallegati2010} and Segarra and Teruel \cite{SegarraTeruel},
show that sampling sizes influence the power-law distribution.

One can fairly assert that the concomitant literature has not yet
reached a consensus regarding what model could best fit empirical data.
An overview of several alternative models is detailed in Ref.
\cite{DeWit2005}, and references therein.
As shown in the above literature review, previous attempts  to
model growth and sizes of firms have not been entirely successful.
In particular, there is a dispute concerning the underlying stochastic
process that steers FSD.
A possible solution in terms of agent-based model was proposed \cite{Farmer2009};
these models are remarkable as descriptive tools,
but they do not furnish an overall panorama because are single-purpose models.
Besides, they are sensitive to the initial conditions,
and, in some cases, their outcome depends on the length of the simulation time.

The aim of this paper is twofold. First,
to develop a thermodynamic-like theoretical model,  able to
capture typical features of firms' distributions. We try to
uncover the putative universal nature of FSD, which could be
characterized by general laws, independent of ``microscopic''
details. Secondly, to validate our theoretical model using an
extensive database of Spanish manufacturing firms during a long
time-period.

This paper contributes to the literature in several aspects.
First, it provides an explanation for the stochastic distribution
of firms' sizes. The understanding of FSD is relevant for economic
policy because it deals with market concentration, and thus, with
competition and antitrust policy measures ---for example, Naldi
\cite{Naldi2003} exhibits a relationship between Zipf's law and
some concentration indices. Second, we apply our model to a large
sample of Spanish firms.  Third, this work  expands the literature
on industrial economics modelling.

The paper is organized as follows.  First, we present the theoretical framework
and perform numerical experiments to validate our analytic approach.
Then, we show the empirical application to the Spanish firms. 
Finally, we draw some discussions and conclusions of our work.

\section{Theoretical framework \label{sec:model}}

Our framework is based on two fundamental hypothesis:
\begin{enumerate}
\item a micro-economic dynamical hypothesis for individual firm growth; and
\item using the maximum entropy principle, with dynamical prior information, for describing  macroeconomic equilibrium.
\end{enumerate}

\subsection{Microdynamics}

\begin{figure}[h]
\centering
\includegraphics[width=0.75\textwidth]{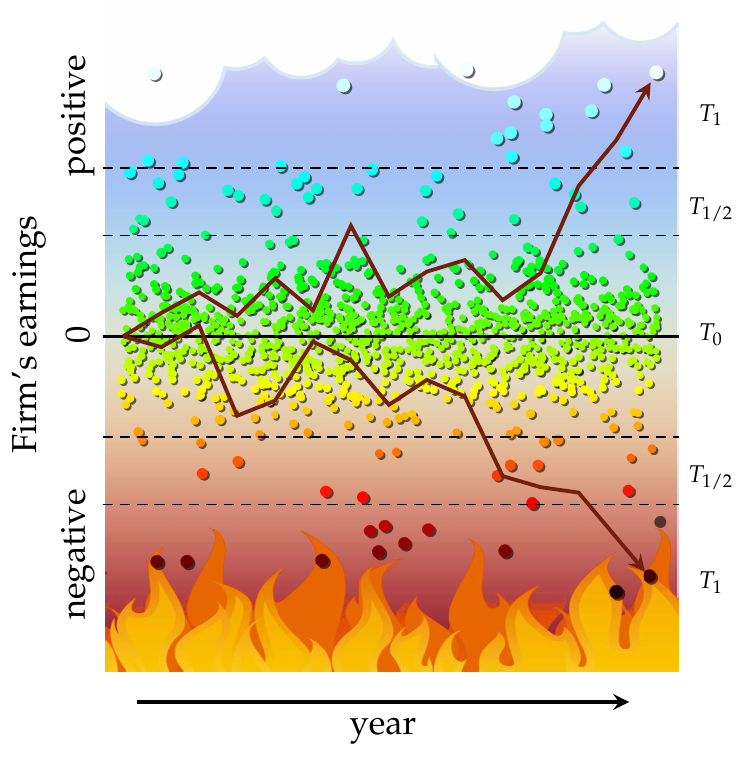}
\caption{Conceptual sketch of firms' dynamics according with Eq. \eqref{eq:eq2} and \eqref{eq:eq3}. 
Firms (represented here by dots) evolve as random walkers in time ---similar to particles in a gas--- according to 
the regime defined by they earnings (delimited by dashed horizontal lines) and the empirical
temperature at that regime. Brown solid arrows show the path for two of the firms.
\label{fig:simple}
}
\end{figure}

For the micro-economical hypothesis, we assume Gibrat's law of proportional
growth (Eq. \eqref{eq:eq1}) as the main mechanism underlying firms' size evolution. 
A finite-size term, due to the central limit theorem \cite{Hernando2013c}, becomes dominant
for medium and small sizes, being proportional to the square root of the size.
In addition to these two terms, we also assume that non-proportional forces
become eventually effective, being dominant for the smallest sizes. Thus, our
full dynamical equation  is written as
\begin{equation}
\label{eq:eq2}
\dot{x}_i(t) = v_{1i}(t) |x_i(t)| + v_{1/2i}(t)|x_i(t)|^{1/2} + v_{0i}(t),
\end{equation}
where $v_{q i}(t)$ ($q=1$, $1/2$ and $0$) are independent growth rates.
It is expected that the growth rates are of a stochastic nature. Thus, a {\it temperature}
can be defined from their variance $T_q = \text{Var}[v_q]$. Accordingly,
assuming that the variation in the growth rates is much larger than the variation in
the observable $x$ ---as done in Refs. \cite{Hernando2012,Hernando2015} --- the variance of the growth for several
realizations becomes
\begin{equation}
\label{eq:eq3}
\text{Var}[\dot{x}] = T_{1} |x|^2 + T_{1/2}|x| + T_{0}.
\end{equation}
This equation defines three regimes, according to the size: small sizes
$x<T_0/T_{1/2}$; medium sizes $T_0/T_{1/2}<x<T_{1/2}/T_1$;
and large sizes $T_{1/2}/T_1<x$. Because of the existence
of the non-proportional term, $x$ is allowed to eventually take negative values. This
defines an additional set of temperatures for the negative domain, which in
principle can be independent of those at the positive one. Since all these
temperatures can be measured from the raw data, their properties can be empirically determined.
In Fig. \ref{fig:simple} we show a conceptual sketch of the ensemble of firms evolving in time as
random walkers along the different regimes, with the corresponding temperature and dynamics for
each of them.

\subsection{MaxEnt Principle}

For an ensemble of firms following
 Eq. \eqref{eq:eq1}, we assume that dynamical equilibrium is asymptotically reached when some
macroscopic constraints
are obeyed. We cite the average total number of firms $N$, the typical wealthiness
of a given particular region, or any other objective observable. In view of the success
of an entropic procedure for describing equilibrium distributions in other social systems
(as, e.g., city population distributions), we take as our macroeconomic hypothesis
the Principle of Maximum Entropy (MaxEnt) with dynamical prior information  \cite{Hernando2012,Hernando2013c,Hernando2012a,Hernando2012b}, to predict the equilibrium
density of the system.
We focus our analytic derivation on that particular  regime that has received most attention
in the literature: the proportional growth-one: $T_{1/2}/T_1<x$ for the largest sizes.
According to Ref. \cite{Hernando2012},
the entropy of a system following Gibrat's law is measured in terms of the new dynamical variable
$u(t)=\log[x(t)/x_c]$ (where $x_c$ is some reference value, in our case, the transition size $x_c=T_{1/2}/T_1$)
which linearizes the dynamical equation as $\dot{u}_i(t) = v_{1i}(t)$. Thus, we write the macroscopic
entropy for the system's density distribution $\rho(u)$ for $N$ firms as
\begin{equation}
S[\rho] = -\int du \rho(u) \log[\rho(u)/N],
\end{equation}
The equilibrium density is obtained by extremization of $S$ under
the empirical constraints \cite{Hernando2012a,Hernando2012b}, such
as the total number of firms, the minimum size of a firms, among
others. Lacking them, as sometimes happens in physics, we will use
a symmetry criterion \cite{CPT87}: employ constraints that
preserve a symmetry of scale of $x(t)$, i.e.  translation symmetry
in $u(t)$. For this, we define an \emph{energy function},
$\mathcal{E}[\rho]$, that depends on powers of the dynamical
variable $u$, namely
\begin{equation}
\mathcal{E}[\rho] = \sum_n\lambda_n\int du \rho(u) (u-\langle u \rangle)^n:= \sum_n\lambda_n m_n,
\end{equation}
where $m_n$ are the central moments of $\rho$ and $\lambda_n$ the coupling constants.
The maximization problem is written as
$\delta_\rho(S[\rho]-\beta\mathcal{E}[\rho]) = 0$, where $\beta$
is a Lagrange multiplier ($\beta\lambda_n$ become then the
multipliers for each term), and the general solution is of the
form
\begin{equation}
\rho(u) = N \exp\left[-1-\beta \sum_n\lambda_n(u-\langle u \rangle)^n\right].
\label{eq:rhoMaxEnt}
\end{equation}
The values of the multipliers are obtained by solving the system
of Lagrange equations, $m_n=\int du \rho(u) (u-\langle u
\rangle)^n$ for the distribution of Eq. \eqref{eq:rhoMaxEnt}.

\subsection{Connection with Thermodynamics}

We consider, for simplicity, the linear regime with only the first
two moments $n=0$ (a constraint on the average total number of firms $m_0=N$) and $n=1$
(a constraint on the mean value $\langle u \rangle$ written as $m_1=0$).
Since the equations are formally equivalent to those found in
Thermodynamics, and traditionally the multipliers associated with
these constraints are \cite{Reif,Balian} $\beta:=1/T_1$,
$\lambda_0:=-\mu$, $\lambda_1:=\lambda$, we have a thermodynamic
potential
\begin{equation}
\Omega = -T_1S - \mu N + \lambda U,
\end{equation}
where $U = \langle u-\langle u \rangle \rangle$. The variational
problem becomes $\delta_\rho \Omega[\rho]=0$. We obtain the
distribution
\begin{equation}
\rho(u) = Ne^{-(\lambda u-\mu)/T_1}.
\end{equation}
The distribution is cast in terms of the observable $x$ as
\begin{equation}
\rho_X(x)dx = Ne^{\mu^*}\frac{x_c^{\lambda^*}}{x^{1+\lambda^*}}dx
\end{equation}
where $\mu^*=\mu/T_1$ and $\lambda^*=\lambda/T_1$. Accordingly, we
obtain a power law density. Useful for analyzing the empirical
data is the complementary of the cumulative distribution
$F(x)=\int^x_{x_c}dx'\rho(x')$, that reads
\begin{equation}
\label{eq:rank}
N-P(x) = N\left(\frac{x_c}x\right)^{\lambda^*}.
\end{equation}
The solutions of the Lagrange equations lead to
\begin{equation}\label{eq:eos0}
e^{-\mu/T_1}=\langle u \rangle;\text{ and }\lambda=T_1/\langle u \rangle,
\end{equation}
and to the {\it equation of state}:
\begin{equation}\label{eq:eos}
e^{\mu^*}=\lambda^*.
\end{equation}
This is the relevant equation for interpreting the empirical data, since $\lambda^*$
can be measured from the data and $\mu^*$ can be interpreted thanks to the thermodynamic analogy.

Indeed, comparing our results with those of a physical system, one
can identify $\mu$ with the chemical potential. We interpret $\mu$
as the ``cost'' for including/creating or excluding/extinguishing
firms in the proportional large-size regime. Following MaxEnt \cite{Hernando2012,Hernando2013c,Hernando2012a}, the
system is in contact with a reservoir of firms, and
 tends to minimize $\Omega$. Since
$\partial\Omega/\partial N=-\mu$, for $\mu>0$ any new firm, in the proportional regime,
decreases $\Omega$, making it more likely the emergence of a flow of firms entering into the system.
However, for $\mu<0$, any new firm will increase the value of $\Omega$, allowing
for a flow of firms exiting the system. In the particular case $\mu=0$ there is no cost
for the flow of firms, in what we expect to be an equilibrium,
stable, and healthy situation for a capitalist economy.

The thermodynamic variable $\lambda$ defines the exponent of the distribution,
and can be interpreted as a measure of the typical wealth of a region.
Specifically, it determines the scale of the size of firms, because it constraints
the geometric mean of $x$. Indeed, the use of the geometric mean instead
of the mean is common for systems with scale invariance, where long-tailed distributions
have undefined moments but well-defined log-moments \cite{IHD,RPIJ_index}. This value will change from
one economy to  other one.

Thanks to the equation of state Eq. \eqref{eq:eos}, we can provide
an intuitive, physically-based interpretation of that exponent:
\begin{itemize}
\item for $\lambda<1$ ($\mu<0$) the system favors the extinction of firms;
\item for $\lambda>1$ ($\mu>0$) the system favors the creation of firms;
\item for $\lambda=1$ ($\mu=0$) the system freely creates and extinguishes firms.
\end{itemize}
This last particular case corresponds to the Zipf's law distribution,
namely
\begin{equation}
\label{eq:Zipf-1}
N-P(x) = N\frac{x_c}{x}.
\end{equation}

\section{Numerical experiments}\label{sec:numerical}

\begin{figure}[h]
\centering
\includegraphics[width=0.75\textwidth]{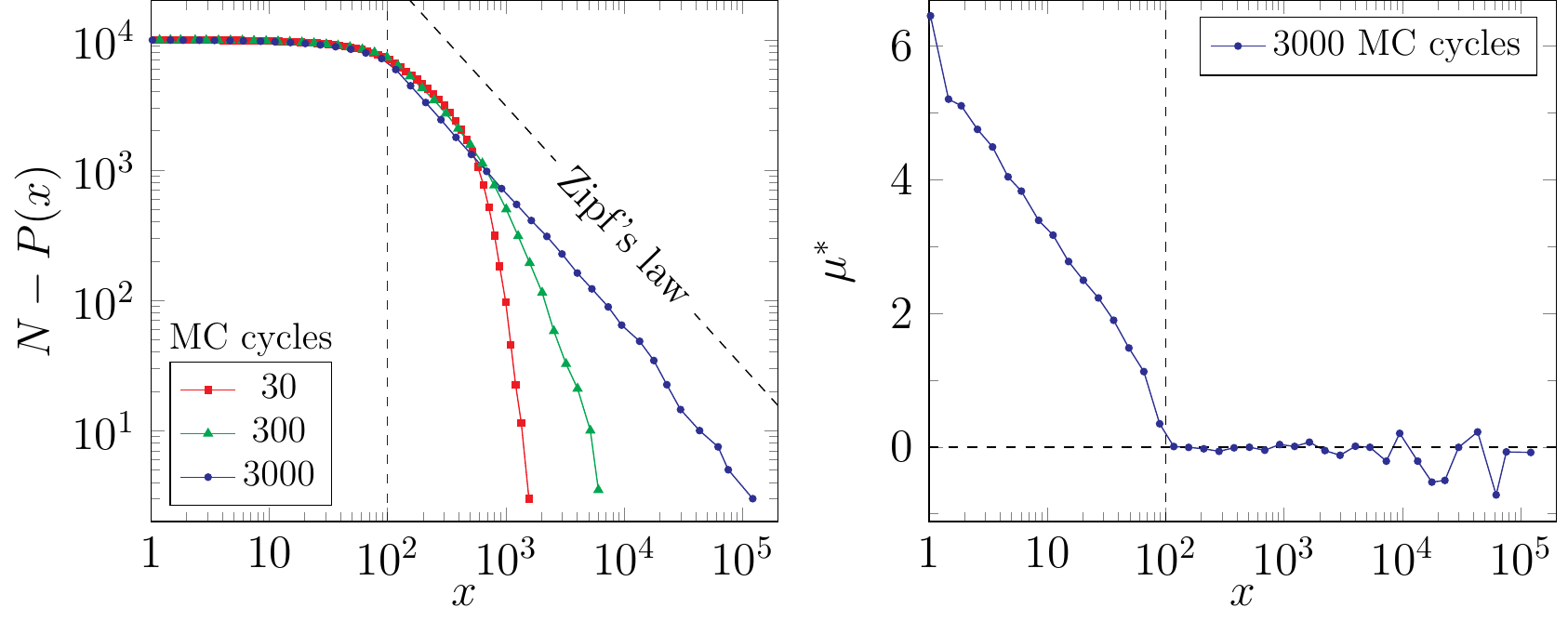}
\caption{Rank plot ({\it left}) and chemical potential ({\it
right}) for the simplified dynamics of Eq. \eqref{eq:simulations}.
We use $N=10^4$ random walkers with transition to proportional growth at 
$x_c=100$ (dashed vertical line). The equilibrium distribution is
reached after $\sim3000$ MC cycles. \label{fig:sim_rank} }
\end{figure}

In aim of testing our theoretical procedure we have performed numerical experiments
in terms of random walkers via a Monte Carlo simulation (MC). At
the initial time, the $N$ random walkers are randomly located
using a uniform distribution. We assume independent stochastic
Wiener coefficients for different firms, within each of the three
regimes or, more explicitly,
\begin{equation}
\langle v_{qi}(t)v_{q'j}(t')\rangle = T_q\delta_{ij}\delta_{qq'}\delta(t-t'),
\end{equation}
where $q$ defines the specific dynamical regime, as in Eq.
\eqref{eq:eq2}. In order to make explicit the mechanisms that
govern the dynamics, we will use a reduced approach, in which each
of the regimes, according to the size $x$, evolves independently.
Therefore, instead of simulating the whole dynamics, i.e. Eq.
\eqref{eq:eq2}, we aim to understand the particular contribution
of each term in that equation. Henceforth, we will focus in the
interplay between the linear and proportional growth regimes,
disregarding the intermediary regime. To achieve this goal, we use
the following equation for the microscopic dynamics:
\begin{equation}
\label{eq:simulations}
\dot{x}_i(t)=
\begin{cases}
v_{0,i}(t),&\textrm{for }x<x_c,\\
v_{1,i}(t)x_i(t),&\textrm{for }x>x_c,
\end{cases}
\end{equation}
where $x_c$ defines the border between the linear and the
proportional regimes. 
Since the number of walkers in  the proportional
regime is not constrained, we follow here the given recipe for
a Grand canonical ensemble \cite{Reif,Balian} 
where $\mu$ is fixed and the fluctuation in the number of walkers is determined by 
the probabilities of including ($P_+$) or extracting ($P_-$) a walker as
\begin{equation}
P_+ \propto e^{\mu^*}\quad\quad
\text{ and }\quad\quad 
P_- \propto e^{-\mu^*}.
\end{equation}
According to these probabilities, in an ensemble with $\mu^*=0$ any walker
can leave or enter into the system without any restriction. Additionally,
following Eq. \eqref{eq:eos0} the constraint $\langle u\rangle=\langle \ln(x/x_c)\rangle=1$
should be fulfilled.
We have performed several
realizations with different initial conditions, and let the system
 evolve until reaching equilibrium. In Fig. \ref{fig:sim_rank}, we
show the rank plot, Eq. \eqref{eq:rank}, for different simulation
times, measured in MC steps. Here we choose $x_c=100$. We see that
the equilibrium distribution (for $\sim 3000$ MC cycles) follows the
Zipf's law. Moreover, we show in Fig. \ref{fig:sim_rank} the
chemical potential for the equilibrium distribution. We see that,
up to some fluctuations, the constraint $\mu^\ast=0$ is respected.

We find that the equilibrium distribution does not depend on the
initial conditions, and follows  Zipf's law: for large values, the
complementary of the cumulative distribution follows Eq.
\eqref{eq:Zipf-1}, as predicted by our thermodynamic framework.
The distribution deviates from the analytic result, Eq.
\eqref{eq:Zipf-1}, as the size of $x$ reaches the transition
critical valued $x_c$.
In view of these results, we numerically
validate our analytic procedure.

\section{Empirical application}\label{sec:empirical}

In order to empirically verify our theoretical model, we consider
the Spanish SABI database
\cite{Dijk2001,SABI}, which is a comprehensive one for all firms
that have the obligation to disclose balance sheets in the Spanish
Mercantile Register. Our sample consists of $1\,155\,142$ firms along 
a decade, with more than $500\,000$ firms per year. We select those 
firms which have been active at any time
during the last 10 years and use as our observable  $x_i(t)$ for
the $i$th firm at year $t$ the so-called Earnings Before Interest,
Taxes, Depreciation, and Amortization (EBITDA). This quantity is
widely employed for assessing companies' performances. It is
homogeneous across companies and is not affected by different
forms of financing. We believe that our proxy for size is a clear
indicator of both corporate performance and size.

\subsection{Microdynamics}

\begin{figure}[h]
\centering
\includegraphics[width=\textwidth]{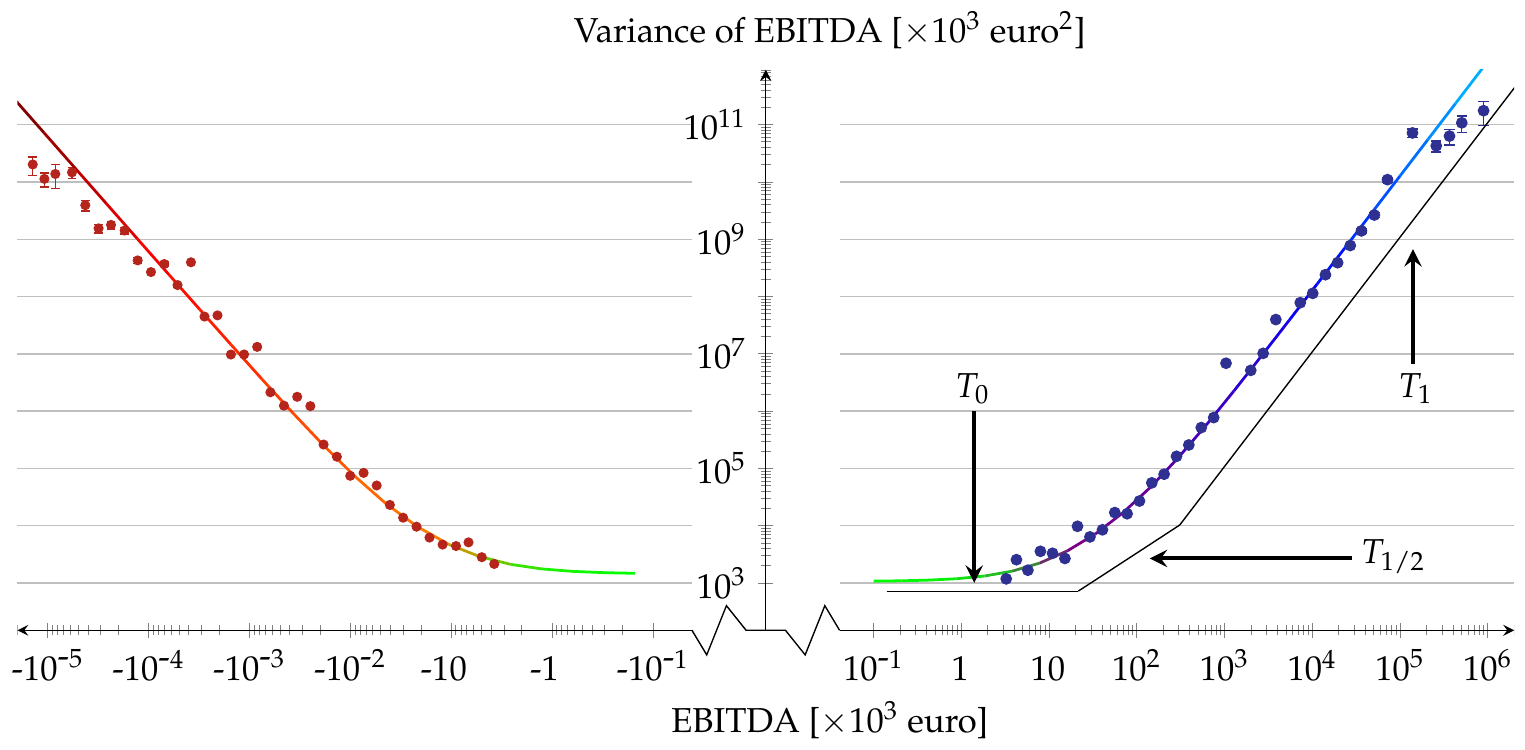}
\caption{Variance of EBITDA for the set of all  Spanish
firms in 2009. The colored lines correspond to the fit according to  Eq.
\eqref{eq:eq2}. The black solid line sketchs the trend-regime:
linear for small firm-sizes, FSC for medium, and
proportional growth for the largest sizes. } \label{fig:spain}
\end{figure}

\begin{figure}[h]
\centering
\includegraphics[width=\textwidth]{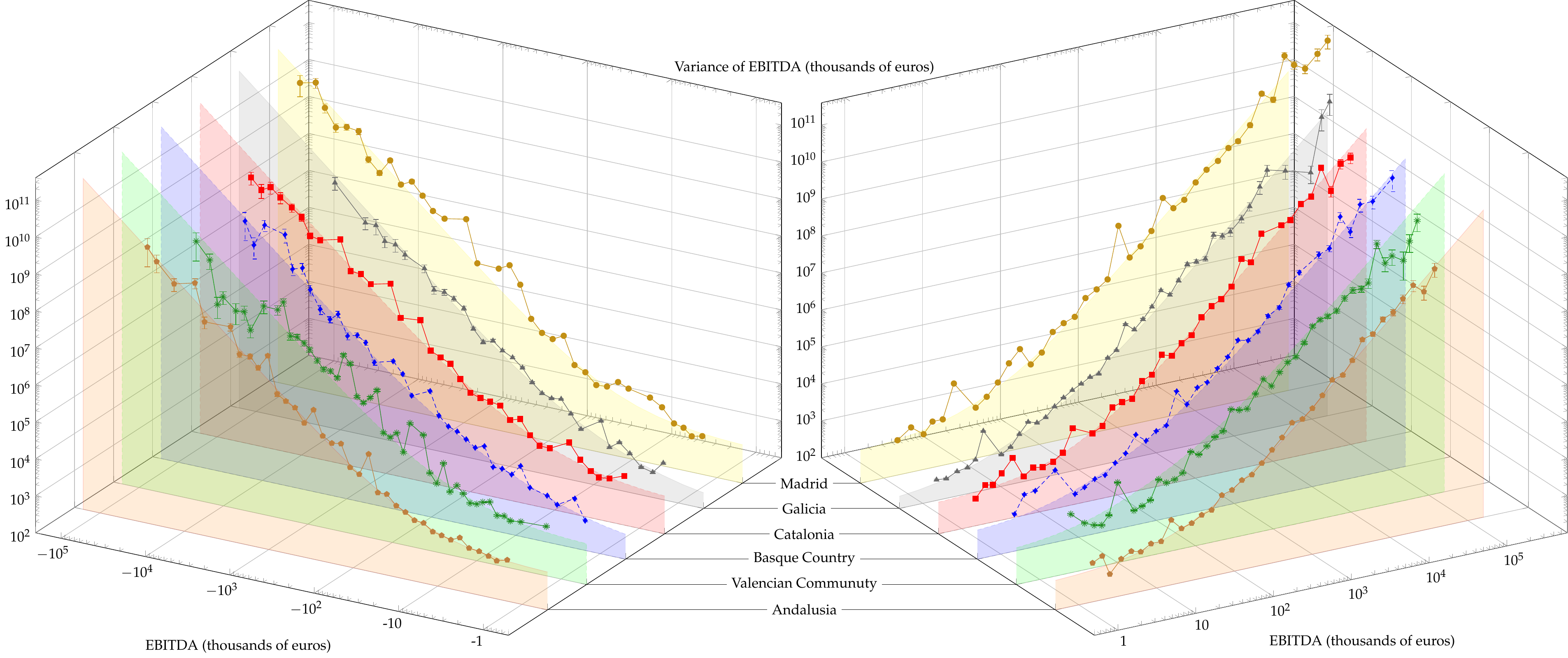}
\caption{ Variance of EBITDA for some Spanish autonomous
communities in 2009. The colored-surfaces correspond to the fit
according to  Eq. \eqref{eq:eq2}. } \label{fig:communities}
\end{figure}

\begin{figure}[h]
\centering
\includegraphics[width=\textwidth]{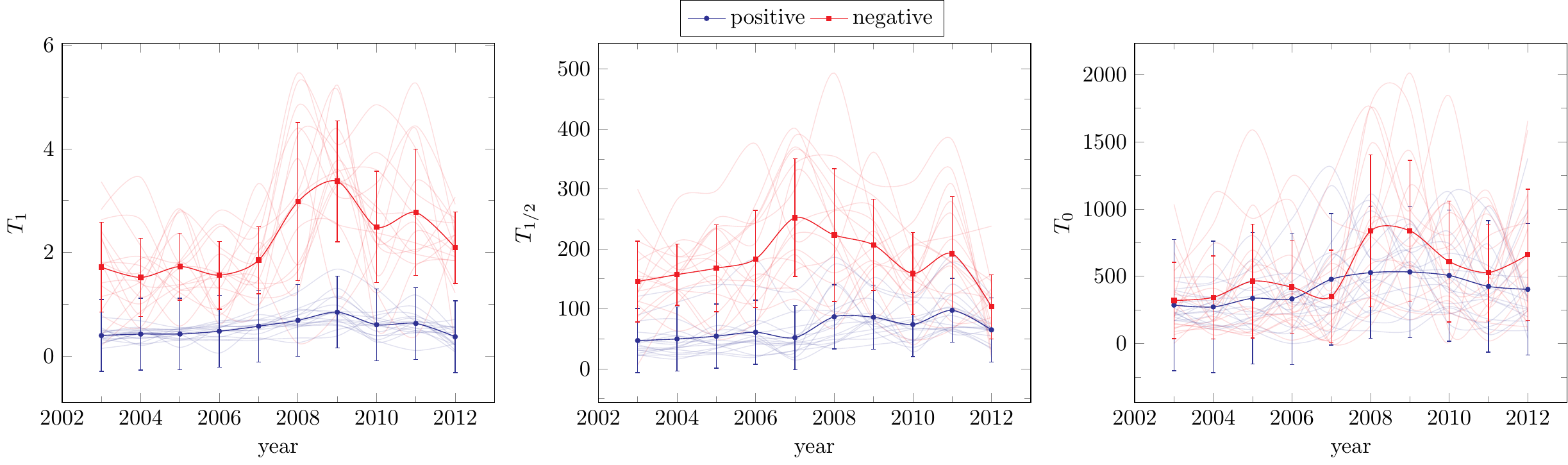}
\caption{Temperatures $T_1$, $T_{1/2}$, and $T_0$ as function of the year.
Faded lines represent the temperature evolution for every autonomous community, bold blue and red lines 
represent the Spanish mean temperature.
\label{fig:temperatures}
}
\end{figure}

We first test our microscopic dynamical hypothesis by measuring
the variance of the EBITDA growth for each year. Since the EBITDA
values can be negative, we study separately positive and negative
domains. We first analyze all the Spanish firms in the same set,
displaying in Fig.~\ref{fig:spain} 
the dependence of the
growth-variance on the EBITDA for the year 2009.  We find a
remarkable match to Eq. \eqref{eq:eq2} for both positive and
negative domains. The transition to proportional growth takes
place, in this case, at $x=T_{1/2}/T_1=109\times10^3$ euro.
Additionally, in both domains, the linear regime temperature,
$T_0$, is of the same order of magnitude. As shown
in the Supporting Information, all the available data for 10 years
match Eq. \eqref{eq:eq2}, with slightly changing temperatures. A
similar analysis, made per each Spanish autonomous community,
shows that the dynamics is also obeyed individually by regions,
as shown in Fig.~\ref{fig:communities}.
We do not find any
exception for all the 15 Spanish autonomous communities during
these 10 years. In view of these results, we empirically confirm
the validity of the dynamical equation \eqref{eq:eq2}.
Additionally, we find that the temperatures $T_1$ and $T_{1/2}$ for
the negative domain are significantly higher than those for the
positive one. Remarkable, 
as shown in Fig. \ref{fig:temperatures}, 
$T_0$ can be considered the same for
positive and negative EBITDA, indicating that the same
non-proportional regime is connecting both domains.

\subsection{Macroequilibrium}

\begin{figure}[h]
\centering
\includegraphics[width=\textwidth]{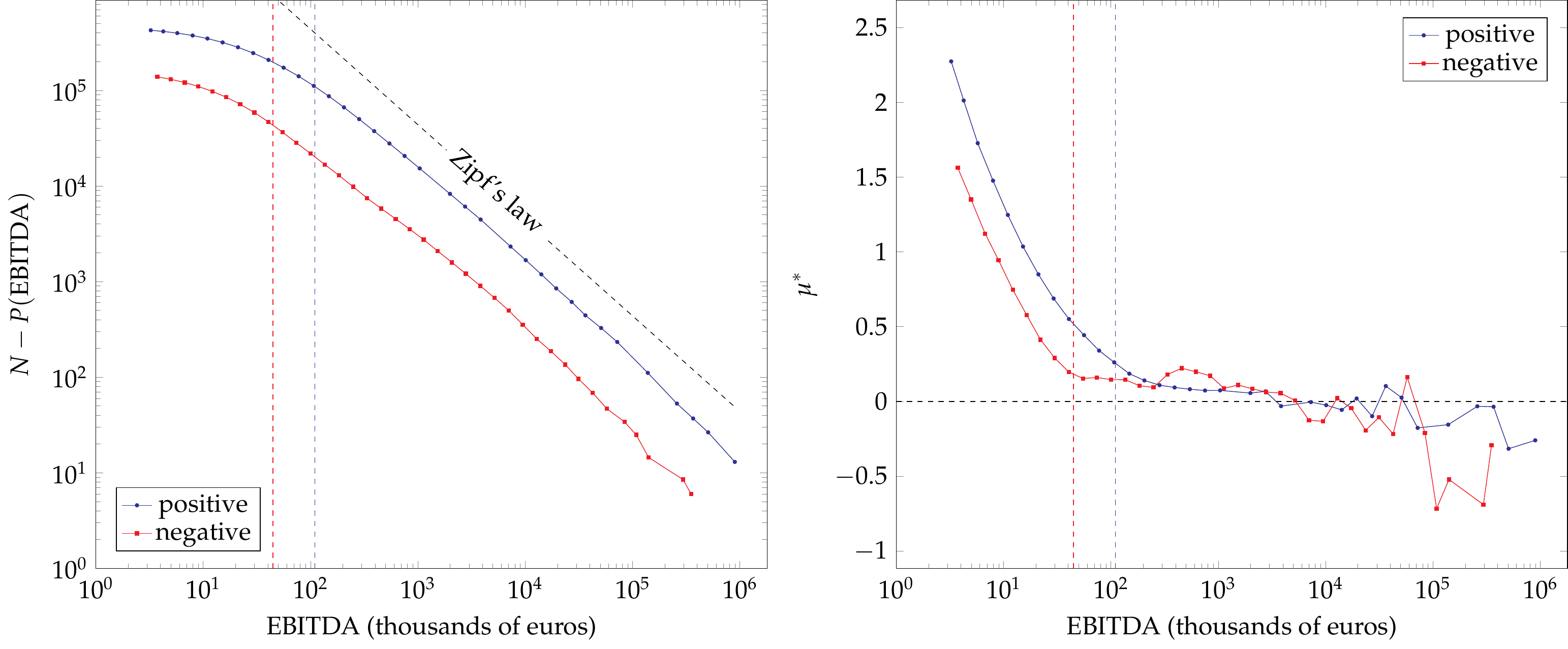}
\caption{Rank plot ({\it left}) and chemical potential ({\it right}) for the empirical
data of Spain in 2009. The observed temperatures for positive EBITDA are
$T_1=1.27\pm0.15$, $T_{1/2}=140\pm20$, and $T_0=1100\pm200$,
and for negative EBITDA $T_1=6.1\pm0.9$, $T_{1/2}=250\pm50$, and $T_0=1500\pm500$.
 The vertical dashed lines display the transition to proportional growth regimes, where the
equation of state \eqref{eq:eos} holds.
\label{fig:spain_rank}
}
\end{figure}

Once the dynamical equation has been validated, we pass to the
rank-distributions. We plot in Fig.~\ref{fig:spain_rank} the
complementary cumulative function $N-P(x)$ for all Spanish firms
in 2009, including those with positive and negative EBITDA. We
observe that for large  values of this function the power law Eq.
\eqref{eq:Zipf-1} is followed, as predicted by our thermodynamic
equilibrium hypothesis, with an exponent very close to that of
Zipf's law $\lambda=1$. For smaller values, the distribution
deviates from the power law. We have checked that this deviation
takes place at about the same transition value $x_c=T_{1/2}/T_1$, as
predicted by our numerical experiments. This is a compelling
evidence for the relation between the dynamics and the
distribution given by our theoretical framework. We also measure
the chemical potential $\mu^*$, via the equation of state Eq.
\eqref{eq:eos}. This chemical potential is, in general, with some
deviations, close to $\mu=0$, value for which the
creation/extinction of firms has no cost for the energy potential
function $\mathcal{E}$. We find the same picture when studying the
firms' distribution per each community. In general, all
distributions are very close to the Zipf's regime. In view of
these results, we consider that our theoretical framework properly
describes the dynamics and equilibrium of the ensemble of Spanish
firms.

\section{Discussion}\label{sec:discussion}

Some interesting assertions can be made with regards to our
theoretical framework. The most relevant, the thermodynamic
interpretation of the exponent $\lambda^*$ of the long-tail in size distribution.
Thanks to the equation of state Eq. \eqref{eq:eos} we provide for the first
time a clear explanation of this exponent, linking that adimensional
number with a dynamical, intuitive mechanism as the cost to the system
of creating or extinguish a firm, measured by the chemical potential $\mu^*$. 
This interpretation can be used to measure
the macroscopic effect of particular economical policies, and to measure
how healthy is a capitalist-based economy. We find that, in general,
the value of $\mu^*$ in Spanish regions is close to zero 
---as shown in Fig. \ref{fig:spain_rank}--- indicating
the freedom of creating or extinguishing a firm.

Additionally, we settle here the form of the microscopic 
dynamics by Eq. \eqref{eq:eq2}, and its dependence
with the size. Contrarily to other social systems following proportional growth
\cite{Hernando2012,Hernando2015,Eliazar:2015}, here exists the possibility
of \emph{negative} values. This requires an additional dynamical
mechanism for the evolution of firms, that is successfully
included in our current approach as a linear term dominant for small sizes.
Firms can be classified according to the dynamical regime,
even if they are in the negative (losses) or positive (gain)
domain. In a pictorial way,  we can talk about
\emph{heaven} (positive proportional regime), \emph{hell}
(negative proportional regime), or \emph{purgatory} (linear
regime). The fact that the temperatures in the positive domain are
systematically smaller than in the negative one, 
as illustrated in Fig. \ref{fig:temperatures}, 
can be summarized
as \emph{hell is warmer than heaven}.
Thus, a firm in {\it hell} losses money in a faster fashion than
it would equivalently earn it in {\it heaven}.

\begin{figure}[h]
\centering
\includegraphics[width=.75\textwidth]{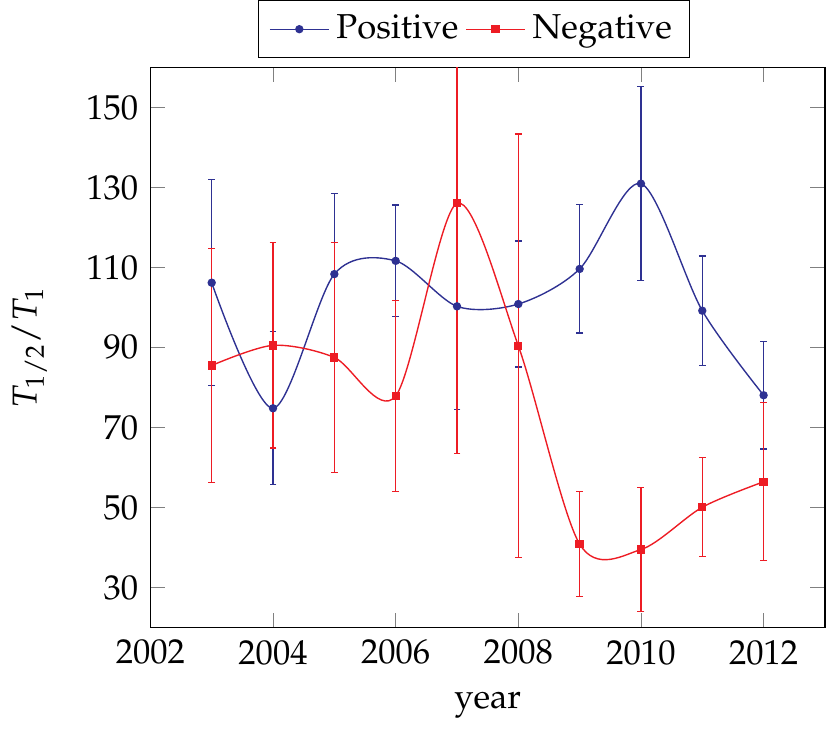}
\caption{Spain's annual transition to the proportional regime.
This is measured by the rate $x_c=T_{1/2}/T_1$. \label{fig:Transition}
}
\end{figure}

We also find useful as macroeconomic indicator the position
of the transition zone between medium and proportional growth
regime in the negative domain, that gives an estimate of the
minimum losses a firm can afford before going bankrupt ---or
metaphorically, the \emph{hell's gate}. Similarly, the same
transition but in the positive domain provides an
estimation of the success region for firms ---that we might
wish to call \emph{heaven's door}. Fig.
\ref{fig:Transition} 
shows both transition values from 2003 to 2012
as measured by the respective temperature ratios $T_{1/2}/T_1$. 
We observe that, before the 2008
Global Financial Crisis, both transitions were approximately
equivalent in size, exhibiting a
symmetry between positive and negative regimes. Right before this
crisis, the negative value reached its maximum, indicating some abnormal 
economic growth, potentially related with 
the speculative bubble. The confidence interval for this specific year is higher
than the absolute value, indicating that this phenomenon did
not happen with the same intensity among all the autonomous
communities. Finally, in the succeeding years, the negative value
was reduced to a half, augmenting the probability of firms to go
bankrupt; whereas the positive transition also decreased,
although not as rapidly as in the negative domain.
After the burst of the crisis, both transitions tend to converge again to 
a similar value, but lower than before the crisis.
Because the equation of state Eq. \eqref{eq:eos} and the constrained
value of $\langle u \rangle$, this lower value reflects a general reduction of the 
wealth in the whole system, because it diminishes, on average, the 
scale of the successful firms at the proportional regime.

As a final remark, the numerical simulation provided here based on
walkers under the Grand canonical ensemble opens the possibility of
developing simulation tools where economical forces can be introduced
in the same fashion as done for physical forces for gases and liquids. Indeed,
we open a bridge between the mathematical tools used
in statistical mechanics and firms' dynamics. Our analytic and numerical
procedures can be used to deeply analyze the empirical data measuring and
parameterizing the economic forces in play, and develop a full quantitative
theory about these dynamics. Work in this line is in progress.

\section{Conclusions}\label{sec:conclusions}

We advanced in this paper a complete thermodynamic framework which
accommodates  the firm size distribution of a given region. We
propose an empirically proof a microscopic dynamical hypothesis,
and show how the firms obey the maximum entropy principle at 
the macroscopic level. We analytically proof the connection between
microscopic dynamics and equilibrium firm's size-distributions
via MaxEnt, and formulate the equation of state that relates
the exponent of the long tail in size distributions with a well-known
thermodynamic observable as the chemical potential.
This lead to a clear and intuitive interpretation of the exponents,
showing that can be used to measure the health of a 
economy. Indeed, the emergence of Zipf's law is associated with the
free cost to the system of creating and extincting firms, as expected
in a capitalist-based economy.
All these theoretical considerations have been validated by comparison
with empirical data concerning Spanish firms, in a window of a
decade. We expect
this work to be a first step towards the formalization of a
theory of the evolution of firms that will yield underlying forces
and laws of evolution.

\bibliographystyle{aipnum4-1}

\newpage

\newpage

\newpage

\newpage

\newpage

\newpage

\newpage

\newpage

\end{document}